\def\year{2023}\relax

\documentclass[letterpaper]{article} 
\usepackage{aaai22}  
\usepackage{times}  
\usepackage{helvet}  
\usepackage{courier}  
\usepackage[hyphens]{url}  
\usepackage{graphicx} 
\usepackage{natbib}  
\usepackage{caption} 
\DeclareCaptionStyle{ruled}{labelfont=normalfont,labelsep=colon,strut=off} 
\usepackage{algorithm}
\usepackage{algorithmic}

\usepackage{newfloat}
\usepackage{listings}
\floatstyle{ruled}
\newfloat{listing}{tb}{lst}{}
\floatname{listing}{Listing}

\usepackage{subcaption} 
\nocopyright 

\usepackage{xcolor, soul}

\title{Examining Similar and Ideologically Correlated Imagery in Online Political Communication}
\author{
   Amogh Joshi and Cody Buntain\textsuperscript{\rm 1} \\
}

\affiliations{
   \textsuperscript{\rm 1}University of Maryland, College Park\\
   College Park, MD 20742\\
   joshi.amoghn@gmail.com, cbuntain@umd.edu
}

\renewcommand\hl[1]{#1}

\begin{document}

\maketitle

\begin{abstract}
This paper investigates visual media shared by US national politicians on Twitter, how a politician's variety of image types shared reflects their political position, and identifies a hazard in using standard methods for image characterization in this context.
While past work has yielded valuable results on politicians' use of imagery in social media, that work has focused primarily on photographic media, which may be insufficient given the variety of visual media shared in such spaces (e.g., infographics, illustrations, or memes).
Leveraging multiple popular, pre-trained, deep-learning models to characterize politicians' visuals, this work uses clustering to identify eight types of visual media shared on Twitter, several of which are not photographic in nature.
Results show individual politicians share a variety of these types, and the distributions of their imagery across these clusters is correlated with their overall ideological position -- e.g., liberal politicians appear to share a larger proportion of infographic-style images, and conservative politicians appear to share more patriotic imagery.
Manual assessment, however, reveals that these image-characterization models often group visually similar images with different semantic meaning into the same clusters, which has implications for how researchers interpret clusters in this space and cluster-based correlations with political ideology.
In particular, collapsing semantic meaning in these pre-trained models may drive null findings on certain clusters of images rather than politicians across the ideological spectrum sharing common types of imagery.
We end this paper with a set of researcher recommendations to prevent such issues.
\end{abstract}

\paragraph{NB} Supplemental material, including the crowdsourcing codebook, is available online: \url{https://osf.io/fkcq8/?view_only=c1c84f72e3bf4ff2a9f37f4ebea4bcbe}

\section{Introduction}

Visual media has long been a key element of political discourse~\citep{Seidman2008,Lilleker2019}, and as new online media spaces increasingly focus on imagery (e.g., Instagram, Snapchat, and TikTok), \hl{new research agendas in computational social science research have emerged, such as images-as-data for political science outlined in} \citet{Joo2018}.
\hl{These new agendas are} facilitated by advances in computer vision and the availability of pre-trained deep learning models for image \hl{analysis}, which has allowed for large-scale, automated characterization of images and their use in online political discourse.
\hl{Common practice for studying political imagery using these computer vision models is to select a pre-trained model and either use it to generate image ``embeddings''--i.e., dense numeric vectors--that characterize images (as in} \citealt{10.1177/1940161220964769.2021} \hl{or} \citealt{10.5117/ccr2022.1.001.joo.2022}) \hl{or fine-tune the selected model for a particular task (as in} \citealt{10.1609/icwsm.v14i1.7338.2020} or \citealt{10.1177/0081175019860244.2019}).
\hl{Despite the proliferation of readily available pre-trained models from repositories like TensorFlow Hub, relatively little guidance is available for computational social science scholars on how to select or use these models.}
\hl{For example, while} \citet{10.1177/00491241221082603.2022} \hl{provides valuable insight into image clustering, that work relies on a single model architecture and task, and how researchers should select models as new ones rapidly emerge remains an open question.} 
\hl{Expanding this guidance is therefore necessary, as insufficient consideration can lead to hazards, such as weakened construct validity, model bias, or partisan asymmetry.}

\hl{This work provides this needed guidance for computational social scientists by analyzing nearly half a million images} shared by 656 Twitter accounts belonging to US members of Congress (MoCs) \hl{using five popular pre-trained computer-vision models spanning seven years of development}.
\hl{Extending the literature on political ideology in visual media, we compare the types of political images produced across these computer vision models, examine what these characterizations reveal about ideological lean in visual media, identify potential asymmetries that emerge across the ideological spectrum, and assess construct validity across similar images within image-types.}

\hl{Results show that, despite varying the underlying pre-trained model, we consistently identify approximately eight classes of image shared by these politicians.}
\hl{In each of these image clusterings, politicians share a variety of image content, with the average MoC sharing at least one image from every image cluster.} 
\hl{These image types range} from photographs containing groups of constituents, headshots, photo ops, and meetings to more diverse visual elements like screenshots of bills, infographics, and meme-like imagery.
Consistent with~\citet{10.1609/icwsm.v14i1.7338.2020}, we find that a politician's distribution of image-types shared is indicative of their ideological position (and therefore party affiliation); e.g., conservative politicians appear more likely to share images with patriotic symbols, whereas more liberal politicians appear to share more images of bills, documents, and infographics.
\hl{We then demonstrate superior performance in predicting MoCs' ideological scores from their average image embeddings compared to prior work.}

\hl{Our analysis also reveals important considerations, however: First, while the majority of pre-trained models produce relatively similar classes of images, the InceptionV3 model diverges substantially from the other models, producing adjusted Rand Indices (ARI) one order of magnitude lower other pairs of models.}
\hl{As newer pre-trained models--EfficientNetB1 and ConvNeXt--are much more similar to the earlier VGG- and ResNet-based models, this result suggests care should be taken when selecting pre-trained models, and one should compare results across several potential models.}
\hl{Second, while we demonstrate MoCs' image-sharing behaviors are predictive of their ideological leans, we find within-party ideological placement is more difficult to predict for Republican MoCs than their Democrat counterparts, regardless of the underlying pre-trained model, which has implications for partisan analyses.}
\hl{Lastly, when assessing construct validity for image ``similarity'' across image clusters, we find, for certain image clusters, human assessors and the pre-trained model disagree in whether a pair of images should be considered similar.}
\hl{That is, for certain types of images, manual and automated identification of similar images agree--e.g., screenshots, infographics, and cartoons--but for other types of images, especially photos showing groups of people, Amazon's Mechanical Turk (MTurk) manual assessment was more likely to disagree.}
\hl{This last point has implications for interpreting model results, as politicians may be using images from one of these problematic clusters in different ways, leading to confounded results (e.g., cluster 2 in EfficientNetB1, which is non-significant but has a higher rate of disagreement from MTurk.}

\hl{Following a description of these results, we discuss connections between our findings and broader literature, especially around prior work on images in political discourse outside social media.}
\hl{The paper outlines recommendations for computational social science researchers who use these models and methods to examine online political discourse, including careful consideration of the types of images being shared, varying use of images by politicians of different parties, and comparing image analyses  from multiple pre-trained models.}
\hl{We then close with a discussion of ethical considerations, threats to broader validity of this work, and potential future avenues to extend this research.}

\paragraph{Contributions}
Overall, this work \hl{is meant to support those studying imagery in online political discourse}, as suggested in \citet{Lilleker2019}, \hl{with} implications for political mobilization (e.g.,~\citealt{Casas2019}), polarization (e.g.,~\citealt{Tucker2018}), and manipulation (e.g.,~\citealt{Zannettou2020}).
Specific contributions include:

\begin{itemize}
\item An analysis of images US politicians share on Twitter \hl{across multiple pre-trained computer vision models};
\item An assessment of the types of imagery that correlate with political ideology;
\item The identification of \hl{several potential hazards} in applying \hl{pre-trained} models to characterize political imagery; and
\item A set of recommendations for researchers when using pre-trained image models to study political ideology in visual media.
\end{itemize}


\section{Related Work}

Studies of imagery in political discourse have a substantial history (see, e.g.,~\citealt{Seidman2008}), and new media spaces like social media platforms provide a new communication medium for politicians to engage with their constituents.
Recent surveys of political operatives shows politicians strategically choose how to use these online spaces~\citep{Kreiss2018}, and the role of visual media in these platforms is becoming increasingly prominent: \citet{Auxier2021} \hl{shows} where visually oriented platforms (e.g., YouTube, Instagram, Snapchat, and TikTok) all see a marked increase in popularity.
At the same time, while much of recent work on politics and social media has focused on text and news sharing, the research community is increasingly calling for greater study of imagery in political discourse~\citep{Lilleker2019}--for instance, \citet{Tucker2018} identify visual media in online political disinformation as a crucial gap in the literature.

While studies have examined visual media in online spaces for political discourse--e.g., \citet{Casas2019} shows online imagery to be particularly impactful in political mobilization--more recent scholarship has begun applying computer vision and machine learning models to facilitate analysis of political imagery.
In this vain, \citet{Joo2018} \hl{surveys} automated methods for analyzing imagery and visual content for political science.
Likewise, \citet{10.1609/icwsm.v14i1.7338.2020} uses a pre-trained deep neural network architecture from \citet{resnet} to characterize images shared by US MoCs on Facebook, evaluating what facial expressions and visual components of the image best correspond to party affiliation.
\hl{These efforts tend toward one of two paths, either generating image embeddings from some pre-trained computer vision model} \citep{10.1177/1940161220964769.2021,10.5117/ccr2022.1.001.joo.2022} \hl{or fine-tuning such a model} \citep{10.1609/icwsm.v14i1.7338.2020,10.1177/0081175019860244.2019}, \hl{but which model to use or the implications of such a choice on the complex, real-world information space remain unclear.}

\citet{10.1177/00491241221082603.2022} \hl{provides foundational guidance for these decisions, but much of the work described therein leverages a single model architecture, based on VGG, with some comparison to ResNet.}
\hl{In this context, } \citet{10.1177/00491241221082603.2022} \hl{examines various cluster counts in grouping images, but how these counts might change across model architectures remains an open question.}
\hl{We address these open points by examining} imagery politicians share in our first research question: \textbf{RQ1} -- \hl{To what extent are the} disparate types of imagery politicians share in \hl{consistent across deep-learning models}?

%

\hl{Besides types of content politicians share, substantial work has examined signals of political ideology that emerge from sharing text} \citep{10.1017/s0007123411000160.2012}, \hl{news} \citep{pew:2017:messing}, \hl{and social interaction} \citep{10.1609/icwsm.v5i1.14126.2021}.
\hl{More recently, } \citet{10.1609/icwsm.v14i1.7338.2020} \hl{has extended these studies to image-sharing, and while that work yields valuable insights into politicians use of images, that work actively removes infographics and other non-photograph content, even stating that infographics appear more used by Democrats.}
\hl{We take a different approach here, extending} \citet{10.1609/icwsm.v14i1.7338.2020} \hl{to a larger set of politicians and images (about a 2x increase), and} evaluate \textbf{RQ2} -- \hl{to what degree an MoC's imagery is predictive of their ideological position, across these models}.

\hl{We further this research by extending to multiple \emph{types} of images rather than filtering out non-photographic images.}
\hl{Integrating the full spectrum of images is necessary, especially with the proliferation of the \emph{visual meme} as potent form of political communication, as discussed in} \citet{10.1609/icwsm.v14i1.7287.2020}.
\hl{Such visual memes include photographs with textual overlays, as with President Trump's declaration that ``Sanctions are Coming'' overlayed on a picture of himself} \citep{10.1609/icwsm.v14i1.7287.2020} \hl{to cartoons, drawings, or panels of images, and are known to play a role in manipulating} online communities~\cite{DBLP:journals/corr/abs-1011-3768,Zannettou2018}.
\hl{Alternatively, screenshots of text or posts have also become important non-photographic forms of visual communication, used to avoid censorship, especially during protests, as seen in the CASM dataset} \citep{10.1177/0081175019860244.2019}.
\hl{Even in CASM, where images must first survive a textual relevance classifier, non-photographic images emerge in the data; and in} \citet{10.1177/00491241221082603.2022}, \hl{a VGG16-based classifier trained on ImageNet collapses non-photographic images into three clusters of symbols and other meme-like images-with-text}.
\hl{These works suggests that} deep learning models trained on primarily photographic datasets like ImageNet~\cite{JiaDeng2009}, may perform \hl{unexpectedly, leading to a loss of construct validity with respect to ``image similarity''.}
To assess this potentiality, \hl{similar to CASM}, we \hl{qualitatively assess image similarity within each class of images, leading to} \textbf{RQ3}--\hl{to what degree do manual assessments about similar pairs of images agree with similarity measures from these pre-trained models}?



\section{Methods}

\subsection{Collecting Congressional Twitter Data}

To analyze politicians' social media content, we leverage a directory of social media accounts for MoCs in the 112th through the 116th US Congressional sessions, covering 2011-2021, primarily from the @unitedstates project,\footnote{\url{https://theunitedstates.io/}} an \hl{open data repository about} the US government.
This \hl{project parses MoC} web pages to extract social media identities, including Facebook and Twitter.
While the @unitedstates project generally contains only social media identities for the current congressional session, the project stores its data in a revision control system.
Using this revision history, we extract \hl{data from past} sessions, manually \hl{augmenting and validating} congresspeople's identities from these sessions (\hl{see } Table~\ref{tab:congresspeople_ids} \hl{for statistics)}.
We observe that, by the 116th session, nearly all MoCs have Twitter accounts, but by our collection period in 2022, nearly 100 accounts have changed Twitter handles or have been deactivated altogether.

\begin{table}[htp]
\centering
    \footnotesize
\begin{tabular}{ l r r r }

\hline\\[-1em]
 & \textbf{} & \textbf{Identified Twitter} & \textbf{Active Twitter} \\
 \textbf{Session} &  {\textbf{MoCs}} & \textbf{Accounts} &  {\textbf{Accounts in 2022}} \\ \hline \\[-1em]

$112^{th}$ & 552 & 367 &  333 \\
$113^{th}$ & 553 & 422 & 395 \\
$114^{th}$ & 547 & 448 & 428 \\
$115^{th}$ & 562 & 495 & 453 \\
$116^{th}$ & 550 & 536 & 465 \\ \hline \\[-1em]
                  & & \textbf{Unique} & 656 \\ \hline

\end{tabular}
\caption{Collected US MoCs Twitter accounts by Session. Numbers of congresspeople do not equal seats in the House and Senate because members can be replaced mid-session.}
\label{tab:congresspeople_ids}
\end{table}%

\subsubsection{Twitter Data}
\hl{We} leverage Twitter's v2 API, with academic access \hl{to collect these politicians'} entire timeline of tweets, going back to 2007. 
We have collected data from 656 Twitter accounts, totaling 3,381,028 tweets.
On average, this dataset contains 5,154 tweets per account, with a minimum of 17, max of 41,870, and standard deviation of~$\sigma=4,649$ tweets.
\hl{Regarding images, this data contains an average of 1,401 images per account, with a minimum of 1, maximum of 11,216 and median of 1,111 images.}

\subsubsection{Sampling Images from Congresspeople's Accounts}
For each of account, we randomly sample 1,111 images \hl{to ensure we get the entire set of images for at least half of the politicians in our dataset.}
This sample does not result in exactly 1,111 images per politician, as some politicians share fewer images.
We do not perform additional processing on images following their download, to allow further analysis to focus on only on the original image content shared. 
In total, this dataset contains 486,604 images across these 617 politicians.
We calculate the mean percent of tweets that contain an image as approximately $37\%$ (i.e., more than one-third of each politician's tweets include an image).

\subsection{Characterizing Types of Images}

After extracting this image sample, we turn to \textbf{RQ1} to characterize the types of images MoCs share.

\subsubsection{Feature Extraction}
We represent each image as an embedded feature vector, extracted from a deep learning model.
To generate these embeddings, we select \hl{five} distinct convolutional neural networks, \hl{ranging from older, more lightweight models to more modern ones}: VGG19~\cite{vgg}, ResNet50~\cite{resnet}, InceptionV3~\cite{inception}, EfficientNetB1~\cite{efficientnet}, \hl{and ConvNext}~\cite{convnext}.
Each model has been trained for image classification on the ImageNet~\cite{imagenet} dataset \hl{for object recognition}. 
\hl{Furthermore, each of these architectures, besides the more recent ConvNeXt model, has been used in studies of images in political discourse.}
Differences between neural \hl{architectures} are primarily an increase in layer depth and decrease in number of individual weight parameters.
In particular, VGG19 employs a direct feedforward sequential architecture, ResNet50 uses ``residual connections'' to merge inputs from prior layers, InceptionV3 builds these residual connections into residual \textit{modules} (consisting of multiple residual layers), EfficientNetB1 uses novel dimension scaling to reduce complexity, \hl{and ConvNeXt incorporates the design of a transformer into a traditional ResNet-based convolutional network.}
The Keras Applications\footnote{\url{https://keras.io/api/applications/}} module \hl{provides pre-trained ImageNet weights for each model}.

To generate image embeddings rather than \hl{ImageNet} class labels, we \hl{replace final model} layer \hl{with} a global average pooling layer.
Each image is represented as a vector with dimension $d$--namely 512, 2048, 2048, 1280, and 2048 for VGG19, ResNet50, InceptionV3, EfficientNetB1, \hl{and ConvNeXt} respectively.
We resize images to $256\times256\times3$ for consistency and extract features in batches of 50.

\subsubsection{Identifying Image-Types via Clustering}
To characterize the disparate types of visuals one can share online--\hl{photos, infographics, cartoons}--as well as the general \textit{content} of these types, we apply k-means clustering to group these images' feature vectors.
We train \hl{five} clustering models (one each for the different deep learning architectures) and set maximum iterations to 1,000.

\begin{figure}[h]
    \centering
    \includegraphics[width=0.95\columnwidth]{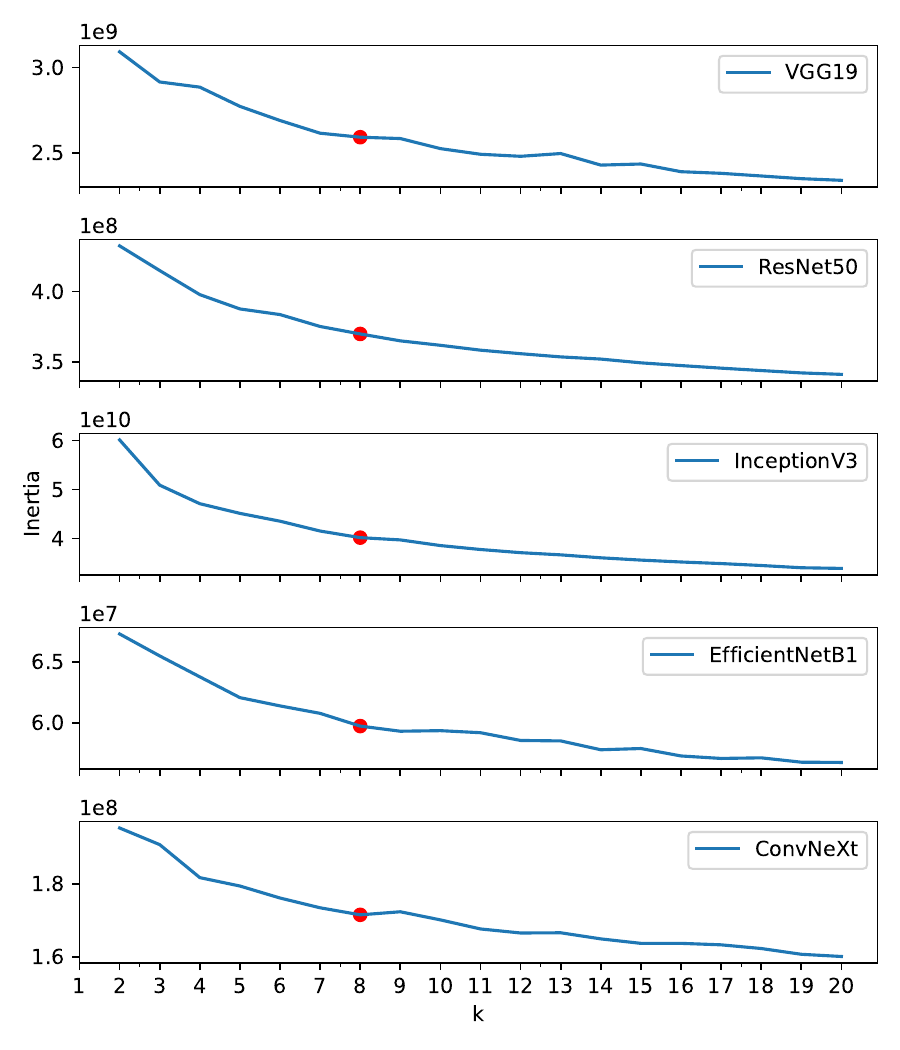}
	\caption{\hl{Inertia Measures per Cluster Count Across Pre-Trained Models. Across all five pre-trained models, we see elbow points at approximately $k=8$.}}
    \label{fig:cluster_metric_diagram}
\end{figure}

To determine clusters $k$, we use the elbow method to identify inflection points in cluster quality metrics, using inertia (sum-of-squares optimization within each cluster).
We plot this metric over $k \in [2, 20]$ and qualitatively determine these inflection points (see Figure~\ref{fig:cluster_metric_diagram}), which appear between $k=4$ and $k=10$, so we settle on $k=8$.
\hl{For robustness, we have compared these results to} the Davies-Bouldin index\cite{4766909} and silhouette scores \cite{9260048}, \hl{finding consistent curves}.

\hl{To assess the similarity across embedding-model clusterings, we use two metrics: First, we calculate the Jaccard similarity between all pairs of clusters and all pairs of models.}
\hl{Second, we use the adjusted Rand Index (ARI) to measure the overall similarity between a pair of clusterings.}
\hl{These metrics provide some insight into how consistent clusterings are across deep-learning models, as needed for \textbf{RQ1}.}

\subsection{Correlating Images and Political Ideology}

For \textbf{RQ2} and correlating ideological position with images, we quantify congresspeople's ideology using their DW-NOMINATE scores~\cite{Lewis2021}, a well-accepted measure of political position \hl{based on voting behavior in Congress}.
\hl{We then use two methods to answer this question: one using clustering to assess which image types reflect ideological lean, and another that predicts MoC ideology directly using their average image embeddings.}

\hl{For our cluster-based analysis, we use} a politician's distribution of images across clusters to predict the politician's ideological position.
This model assigns each politician a numerical value on a left-to-right (i.e., liberal-to-conservative) scale, where negative scores represent stronger liberal positions, and positive scores indicate stronger conservative positions.
We then construct a linear regression model to predict these ideology scores from a politician's distribution of images across clusters.
The primary factors in this model are the proportions of images a politicians shares in each cluster, since politicians may have shared different numbers of images.
To this end, we compile data for the regression model using a slightly reduced set of politicians.
Out of \hl{656} politicians in our dataset, we remove any accounts who have shared fewer than \hl{20} images, or do not have a corresponding DW-NOMINATE score, leaving us with \hl{627} politicians.

\hl{For our second method to predict MoC ideology, we train a set of supervised learning models--one for each neural architecture--to predict DW-NOMINATE scores from an MoC's image embedding, averaged across all images that MoC has shared.}
\hl{Using repeated random sub-sampling across multiple regression models, we estimate the Pearson correlation coefficients between model output and DW-NOMINATE scores.}
\hl{We estimate this metric on the 20\% held-out set of MoCs for each round of 128 random sub-samples across both the full set of MoCs and within the Democratic and Republican parties separately, as in} \citet{10.1609/icwsm.v14i1.7338.2020} and \citet{10.1093/pan/mpu011.2017}.

\subsection{Manually Assessing Image Similarity}

To evaluate \textbf{RQ3} and \hl{assess construct validity of image similarity in the context of these deep neural models}, we leverage MTurk to crowdsource assessments for pairs of ``similar'' images.
We sample similar pairs of images across each cluster and ask three MTurk workers to assess these pairs for their visual and semantic similarities.
\hl{The more MTurk workers agree that pairs of images are similar, the higher the construct validity of image similarity in these embeddings.}

\subsubsection{Sampling Similar Images}

Our primarily objective in this assessment is to understand whether two images that are similar in the embedding space are also considered similar \hl{by a human assessor}; that is, whether an individual would say these two images should in fact be considered similar.
Hence, we extract samples of pairs containing similar images from each of the eight clusters \hl{produced from EfficientNetB1 embeddings, as this model has been used in prior work on visual political discourse}.

For each cluster, we extract \hl{samples of highly similar images.}
Given the scale of this dataset, we cannot compute all possible pairs and instead use locality sensitive hashing (LSH) to find a sample of similar pairs.
\hl{For LSH, we set an image-similarity threshold of} Euclidean distances less than 10. 
\hl{This threshold was chosen based on a pilot study where we sampled pairs of images across the dataset to identify potential duplicates.}
\hl{Below this threshold, several clusters have no potential duplicates for manual assessment, and above this threshold resulted in samples that contained clearly different image-pairs, which would dilute the value of the human assessment.} 
We then take the top 100 most similar pairs per cluster as our sample, creating 800 similar-image pairs.

\subsubsection{Manually Assessing Visual and Semantic Similarity}
To measure how well these clusters actually capture image similarity, we use MTurk to crowdsource similarity assessments for each of these image pairs.
In preliminary analysis, we have found that some pairs of images, such as screenshots of bills or letters, are highly similar visually but contain vastly different messages.
This issue is problematic as clusters of semantically distinct images may exhibit differential use across the ideological spectrum, thereby confounding approaches that rely on visually oriented, automated image analysis.
We have therefore developed a codebook to assess this visual and semantic similarity, available along with the raw data in the online data repository.\footnote{\url{https://osf.io/fkcq8/?view_only=c1c84f72e3bf4ff2a9f37f4ebea4bcbe}}
This codebook describes five categories of image-pairs: identical images; visually and semantically similar images--visually similar but semantically distinct images; visually distinct but semantically similar images; and visually and semantically distinct images--plus a label for ``unknown'', where the assessor cannot determine similarity.

We create a task on the MTurk platform using these pairs with this codebook as instructions for the MTurk workers.
Each image pair receives assessments from three workers, and we take the majority vote as the similarity assessment for that image pair.
In a pilot study, MTurk workers have performed approximately three assessments a minute, so, to pay a minimum wage of \$15 per hour, we pay \$0.10 per assessment.
This crowdsource project has also been reviewed and approved by our university Institutional Review Board.

%



\section{Results}

\begin{figure*}[h!]
    \centering
    \includegraphics[width=0.9\textwidth]{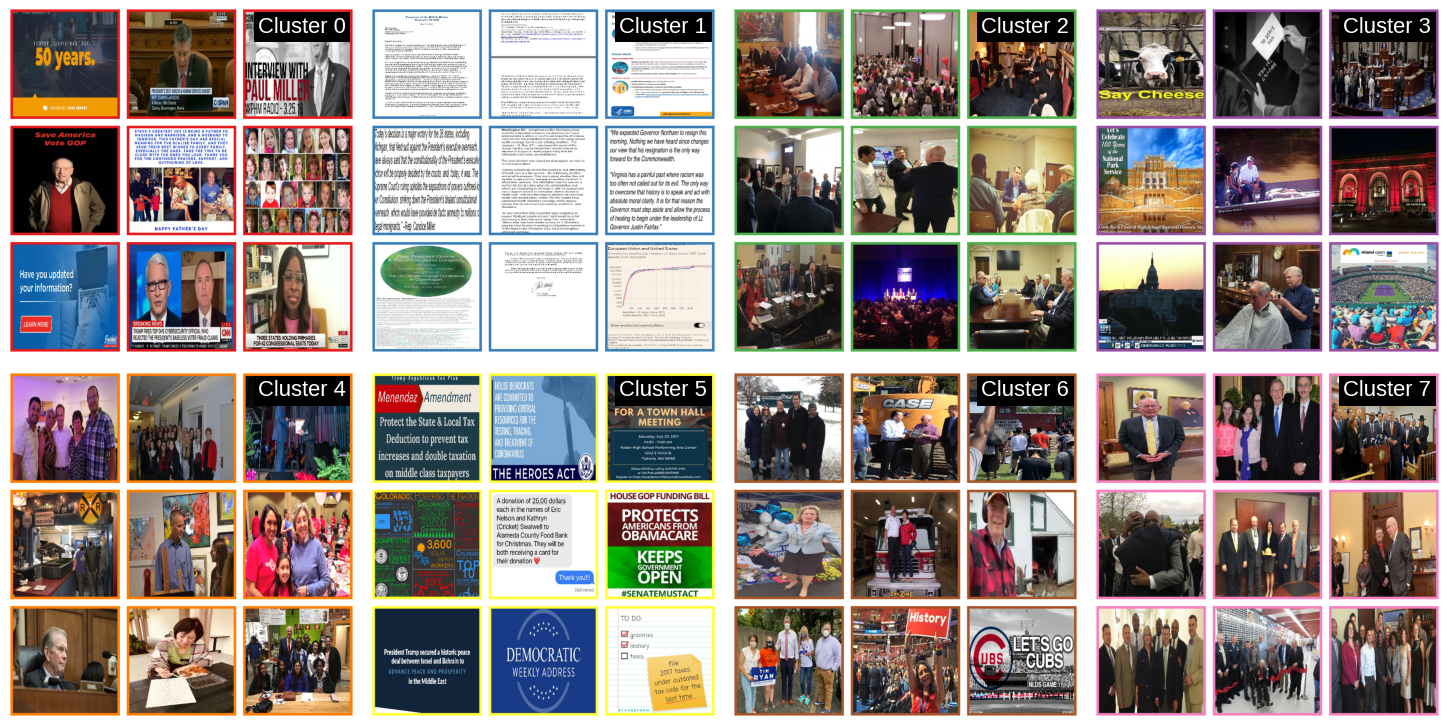}
    \caption{Sample Images from EfficientNetB1 Clustering. Clusters 1, 4, 5, and 6 are correlated with liberal ideology, and Clusters 3 and 7 are correlated with conservative ideology.}
    \label{fig:efficientnet_clusters}
\end{figure*}

\hl{To ground the following analyses,} Figure~\ref{fig:efficientnet_clusters} shows example images from each cluster \hl{in the EfficientNetB1 clustering}.

\subsection{Consistency of Image Types Across Models}

\hl{Examining the frequency of MoCs in each cluster, across all five models, we find that} each cluster contains images from \hl{nearly all} MoCs.
\hl{On average, an MoC shares at least one image from each of the eight clusters, regardless of the underlying deep-learning model.}
\hl{This result shows that MoCs tend to share a variety of imagery as opposed to a single type, regardless of party affiliation and model.}

\begin{figure*}[!t]
    \centering
    \includegraphics[width=\textwidth]{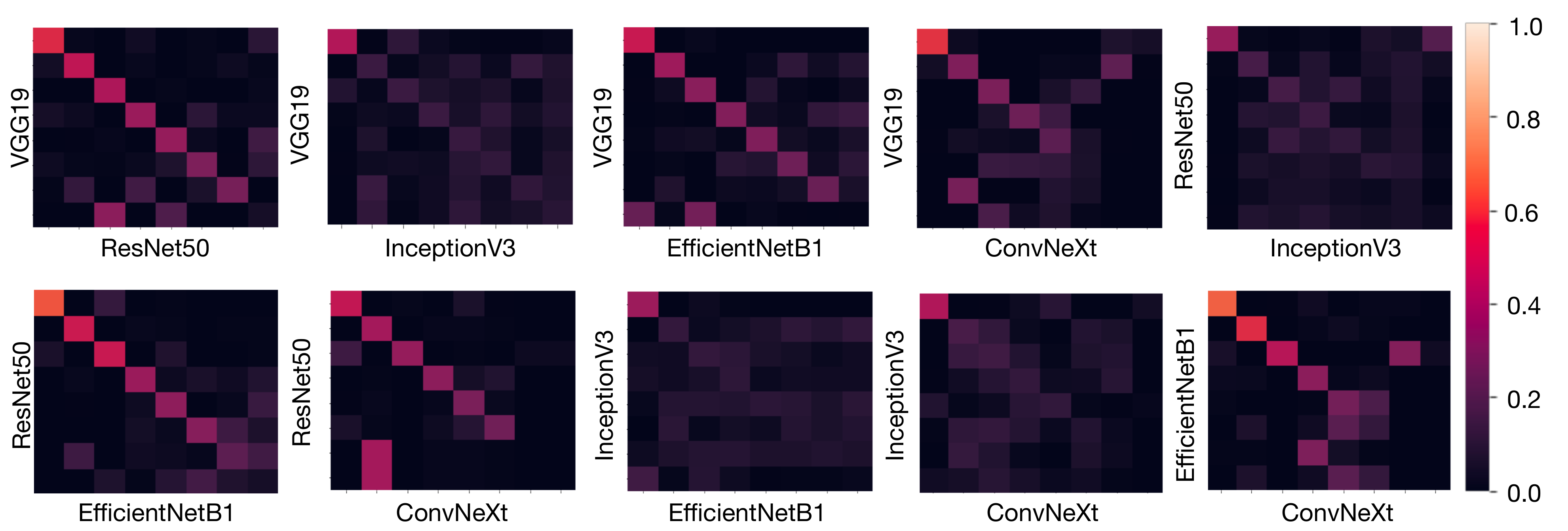}
    \caption{Jaccard Similarity Across Clusters and Embedding Models. \hl{Top-left cells in each grid represent the cluster pair with the highest Jaccard similarity between two models.} VGG19, ResNet50, EfficientNetB1, and ConvNeXt produce relatively similar clusterings, whereas InceptionV3 diverges. \hl{VGG19 and ResNet50 produce the highest $ARI=0.32$, with EfficientNetB1 and ConvNeXt with the second highest, $ARI=0.31$.}}
    \label{fig:jaccard_heatmaps_large}
\end{figure*}

\hl{These findings yield some insight into \textbf{RQ1} in that MoCs' behaviors are largely consistent across deep-learning models, but we still need to assess how consistent the clusterings are across these models.}
\hl{To this end,} Figure~\ref{fig:jaccard_heatmaps_large} \hl{shows aligned clusters for each pair of clusterings.}
This figure shows \hl{VGG19, ResNet50, EfficientNetB1, and ConvNeXt} produce \hl{relatively consistent} clusters, \hl{as each of these pairs have \emph{multiple} pairs of aligned clusters}. 
In contrast, InceptionV3 \hl{consistently produces only a single cluster that is clearly aligned with another model's clusters--e.g., cluster 0 in EfficientNetB1.}
\hl{This examination of pairwise Jaccard similarity is supported by the ARI scores, where all pairs of clusterings excluding the InceptionV3 have an ARI $>0.2$, whereas any pair with InceptionV3 has an ARI $<0.062$, demonstrating that this model produces  divergent results.}

\subsection{Correlating Images and Political Ideology}

\begin{table*}[h!]
    \centering
    \begin{tabular}{l r r r r r r r r r r}
    \hline\\[-1em]
    
         & \multicolumn{2}{c}{\textbf{VGG19}} & \multicolumn{2}{c}{\textbf{ResNet50}} & \multicolumn{2}{c}{\textbf{InceptionV3}} & \multicolumn{2}{c}{\textbf{EfficientNetB1}} & \multicolumn{2}{c}{\textbf{ConvNeXt}} \\\cline{2-3}\cline{4-5}\cline{6-7}\cline{8-9}\cline{10-11}\\[-1em]
         & \textit{Coeff} & \textit{SE} & \textit{Coeff} & \textit{SE}  & \textit{Coeff} & \textit{SE} & \textit{Coeff} & \textit{SE} & \textit{Coeff} & \textit{SE} \\\hline\\[-1em]

        Const      & -0.11$^{***}$   &   0.04     &    -0.01$\;\;\;\;\:$            &   0.02    &     0.05$\;\;\;\;\:$         &   0.09   &    -0.05$\;\;\;\;\:$        &    0.03   &    -0.04$\;\;\;\;\:$        &    0.04  \\
        Cluster 0   &  0.58$\;\;\;\;\:$           &   0.35     &    -0.10$\;\;\;\;\:$            &   0.26    &    -0.74$\;\;\;\;\:$         &   0.46   &    -0.06$\;\;\;\;\:$         &   0.25   &    -0.03$\;\;\;\;\:$        &    0.27  \\
        Cluster 1   &  0.59$^{**}\;\:$    &   0.22     &     0.87$^{**}\;\:$     &   0.29    &     1.04$\;\;\;\;\:$         &   0.59   &    -1.01$^{***}$ &   0.22   &    -1.03$^{***}$        &    0.26  \\
        Cluster 2   & -2.18$^{***}$   &   0.37     &    -0.19$\;\;\;\;\:$            &   0.25    &    -1.38$^{**}\;\:$  &   0.42   &     0.52$\;\;\;\;\:$         &   0.31   &    -3.53$^{***}$        &    0.42  \\
        Cluster 3   & -0.82$^{*}\;\;\:$     &   0.35     &    -0.76$^{**}\;\:$     &   0.26    &     1.64$^{**}\;\:$  &   0.49   &     2.44$^{***}$ &   0.25   &    0.77$^{**}\;\:$        &    0.23  \\
        Cluster 4   &  0.59$^{*}\;\;\:$     &   0.29     &     2.79$^{***}$    &   0.29    &    -2.75$^{***}$ &   0.50   &    -0.67$^{*}\;\;\:$   &   0.33   &    0.71$^{**}\;\:$        &    0.24  \\
        Cluster 5   &  1.37$^{***}$   &   0.26     &    -0.91$^{***}$    &   0.21    &     1.47$^{**}\;\:$  &   0.50   &    -1.28$^{***}$ &   0.24   &    -0.60$^{***}$        &    0.13  \\
        Cluster 6   &  1.34$^{***}$   &   0.28     &     0.25$\;\;\;\;\:$            &   0.28    &     2.77$^{**}\;\:$  &   1.05   &    -0.96$^{***}$ &   0.26   &    0.46$^{**}\;\:$        &    0.17  \\
        Cluster 7   & -1.57$^{***}$   &   0.34     &    -1.96$^{***}$    &   0.38    &    -1.20$^{**}\;\:$  &   0.66   &     0.98$^{***}$ &   0.26   &    3.19$^{***}$        &    0.32 \\

        \hline\\[-1em]
        $R^2$ & \multicolumn{2}{c}{0.19} & \multicolumn{2}{c}{0.20} & \multicolumn{2}{c}{0.12} & \multicolumn{2}{c}{0.23} & \multicolumn{2}{c}{0.29}

\\\hline\\[-1em]
\multicolumn{11}{r}{\textit{Note:} $\;^*p < 0.05$, $^{**}p< 0.01$, $^{***}p< 0.001$}
\\\hline
    \end{tabular}
    \caption{\hl{Linear Regression Correlating Politicians' Cluster Distributions with DW-NOMINATE Scores. The more positive the coefficient, the more conservative-leaning the politician. Results show each embedding model produces five to seven clusters that correlate significantly with ideology.}}
    \label{tab:ols_results}
\end{table*}

Moving to \textbf{RQ2}, we evaluate whether a politician's distribution of images across these eight clusters correlates with that politician's ideology via a linear model that regresses MoC's ideologies on cluster distributions.
Table~\ref{tab:ols_results} shows this linear model fitted across clusterings from the \hl{five} embeddings, demonstrating that an MoC's use of images in particular clusters significant correlates with ideology.
These clusters are fairly balanced across the ideological spectrum as well, as \hl{nearly half of the} clusters correlate with liberal and conservative ideological lean respectively.
In EfficientNetB1, for example, the more an MoC posts images in Cluster 5, the more liberal they are likely to be, whereas the more images they post from Cluster 3, the more conservative their DW-NOMINATE position.
Examining the adjusted $R^2$ for these models, we see that the \hl{linear model for ConvNeXt is the highest, followed by EfficientNetB1, ResNet50, and VGG19 in order}. 
\hl{As in RQ1, InceptionV3 again deviates from this pattern, showing the lowest $R^2$ despite being a newer model than VGG19 and ResNet50}.



\begin{figure}[h!]
    \centering
    \includegraphics[width=0.45\textwidth]{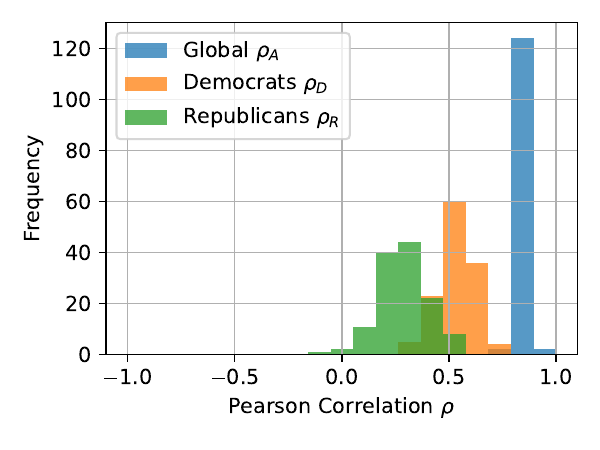}
    \caption{\hl{Overall and Within-Party Correlations Between DW-NOMINATE and EfficientNetB1 Predictions. Overall correlation is strong  ($\rho_A = 0.85$), and within-party correlations for Democrats  ($\rho_D = 0.53$) are substantially higher than for Republicans  ($\rho_A = 0.29$).}}
    \label{fig:direct_ideology}
\end{figure}

\hl{Though these results demonstrate that certain image types consistently correlate with an MoC's  ideology, we also assess the predictive power of an MoC's images.}
\hl{For each MoC, we train a Bayesian ridge regression model to predict DW-NOMINATE scores given an MoC's average image embeddings from EfficientNetB1.}
\hl{Following repeated random sub-sampling to estimate Pearson correlation coefficients, Figure } \ref{fig:direct_ideology} \hl{shows both the global ($\rho_A = 0.85$) and within-party correlation between predicted ideologies and DW-NOMINATE scores, where within-party correlations are separated by MoCs with Democratic ($\rho_D= 0.53$) and Republican ($\rho_R = 0.29$) party affiliations.}
\hl{We make two key observations from these results: First, both overall and within-party correlations outperform those presented in} \citep{10.1609/icwsm.v14i1.7338.2020} \hl{by a wide margin (approximately 26\% overall, 105\% for Democrats, and 25\% for Republicans), with Democrats' image-sharing model within 15\% of the social network-based metric presented in} \citep{10.1093/pan/mpu011.2017}.
\hl{Second, the model consistently underperforms for Republican legislators compared to Democrats, where within-party ideology scores for Democrats are about twice as strong as Republicans.}
\hl{For robustness, we have checked these results across VGG19, ResNet50, and ConvNeXt as well as several regression models, and results are consistent both in comparison to prior work and in the Democrat-Republican asymmetry.}

\begin{figure*}[!t]
    \centering
    \includegraphics[width=0.9\textwidth]{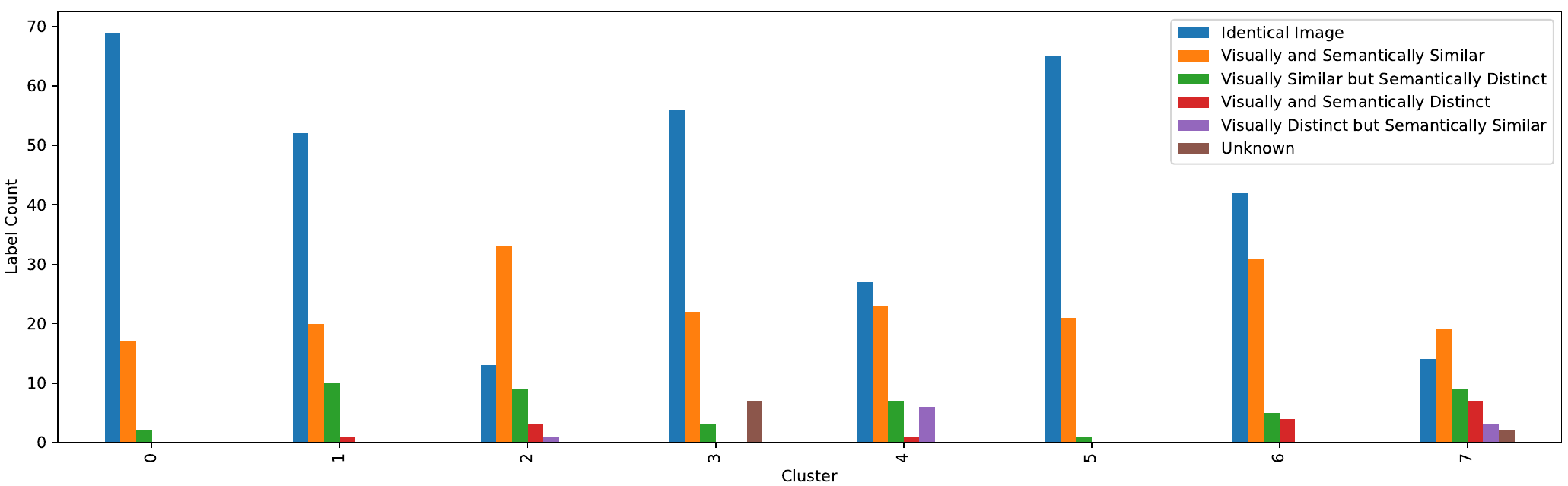}
    \caption{Crowdsourced Majority-Vote Labels of Similarity per EfficientNet B1 Cluster. Clusters 0, 1, 3, and 5 contain mostly identical images; clusters 2, 4, and 6 contain visually similar images; and cluster 7 contains more distinct images.}
    \label{fig:similar_imagery_cluster_proportions}
\end{figure*}


\subsection{Assessing Image Similarity and Construct Validity}

We have shown that MoCs share a variety of images (\textbf{RQ1}) and that \hl{images shared are predictive} of their political ideology (\textbf{RQ2}).
\hl{For \textbf{RQ3}, we investigate whether similarity within a particular image type in embedding space are consistent human assessments--i.e., whether similarity in the metric space matches how an MoC might see image similarity.}
Using embeddings and resulting clusters from EfficientNetB1, we extract a sample of 100 highly similar pairs of images per cluster and use MTurk to collect three manual assessments of image similarity per pair.
Table~\ref{tab:manual_assess_labels} shows label frequencies for raw counts across 384 MTurk workers and the label counts after taking the majority vote and removing 195 samples that lack clear consensus.
\hl{Of the image pairs with a majority vote, MTurk workers say 94\% are visually similar, with 8\% being only visually similar but semantically distinct.}
\hl{This high proportion of agreement between embeddings and human assessment provides confidence in the construct validity of this image similarity metric.}

\begin{table}[htbp]
    \scriptsize
    \centering
    \begin{tabular}{l r r}
    \hline
               & Raw Count  & Majority Vote \\ 
     Label & ($n=2,400$) & ($n=605$) \\     \hline
    
Identical Image & 1,020 & 338 \\
Visually and Semantically Similar & 726 & 186 \\
Visually Similar but Semantically Distinct & 318 & 46 \\
Visually and Semantically Distinct & 172 & 16 \\
Visually Distinct but Semantically Similar & 117 & 10 \\
Unknown & 47 & 9 \\
    
\hline
    \end{tabular}
    \caption{Distributions of MTurk Similarity Assessments. Most image pairs are visually similar, but MTurk workers do not reach consensus in $195$ of these image-pairs.}
    \label{tab:manual_assess_labels}
\end{table}

\hl{Looking within individual clusters, however, a difference emerges.}
Figure~\ref{fig:similar_imagery_cluster_proportions} shows clusters 0, 1, 3, 5, and 6 predominantly contain images that are either identical or visually and semantically similar ($>73\%$).
Clusters 2, 4, and 7, however, have fewer identical image pairs and more pairs without a clear consensus ($36-46\%$), suggesting \hl{that construct validity is weaker for these types of images}.
Of these three EfficientNetB1 clusters, Table \ref{tab:ols_results} shows Cluster 4 leans liberal, and Cluster 7 leans conservative (i.e., has a positive, significant coefficient).
Cluster 4's significance is borderline, however, and Cluster 2 shows no significant predictive power at all.
All three clusters primarily show groups of people in various situations, and similarity assessments may be difficult for humans in these cases.
Consequently, for Cluster 2 (and Cluster 4 to a degree), the absence of significant effect in Table \ref{tab:ols_results} may be attributable to \hl{differences between embeddings and} human perception rather than ideological use.
These results of identical and visually/semantically similar images in a majority of clusters is also consistent with a pilot we have run using the ResNet50 embedding model and manual annotation by two authors.

\section{Discussion}

The results above broadly demonstrate that multiple pre-trained models for image characterization are largely consistent in both the types of imagery they identify and how those types of imagery correlate with politicians' ideological positions.
Despite this consistency, our results also demonstrate potential areas of concern.
Below, we discuss some of these factors, namely about the types of imagery we identify, how they are used across the political spectrum, and potential threats to our results.

\begin{figure*}[h!]
    \centering
    \subcaptionbox{Document-Based Imagery}[8.5cm]{\includegraphics[width=0.85\columnwidth]{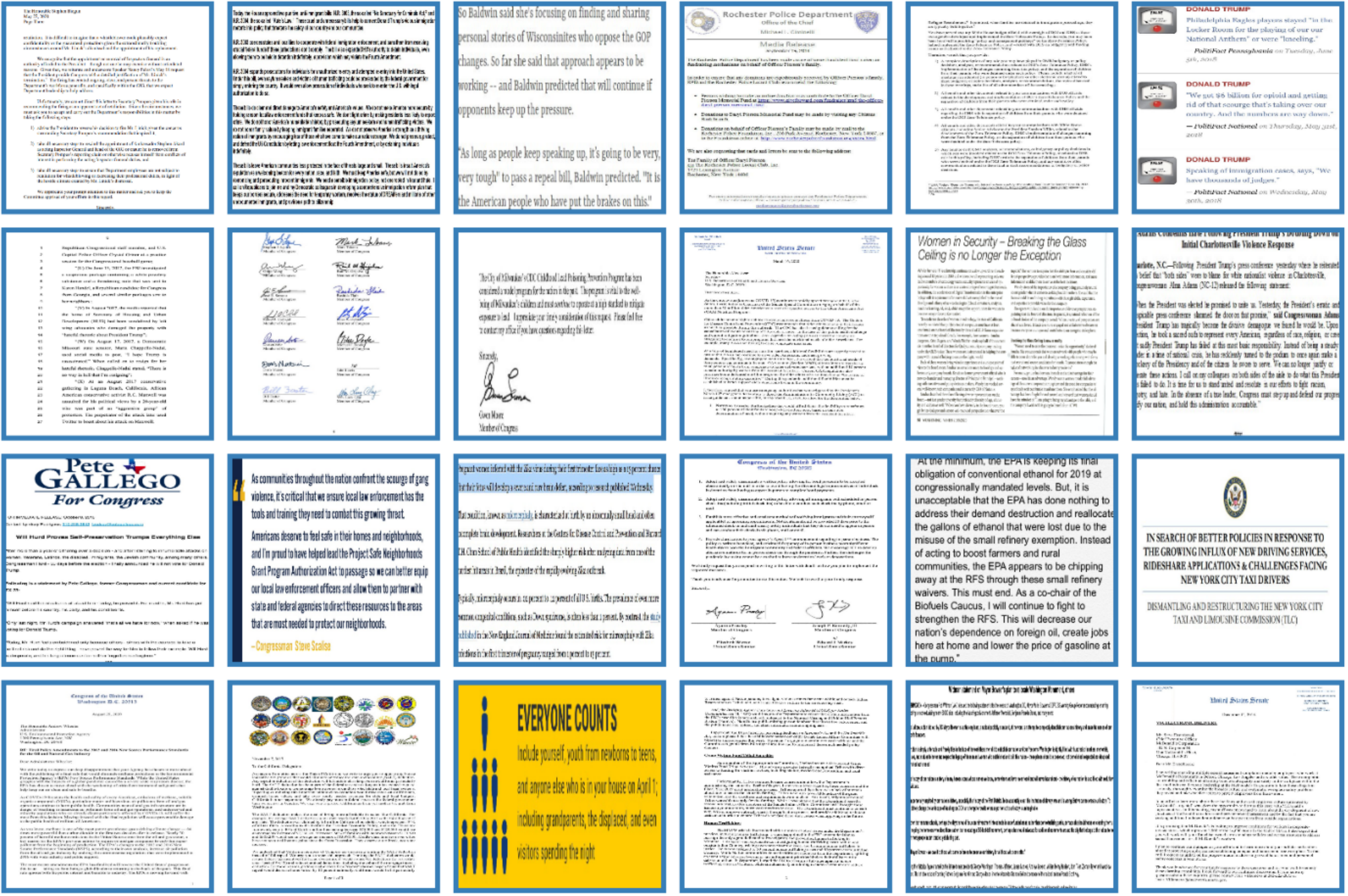}\label{fig:ideological_imagery_types_docs}}
    \subcaptionbox{Patriotic Imagery}[8.5cm]{\includegraphics[width=0.85\columnwidth]{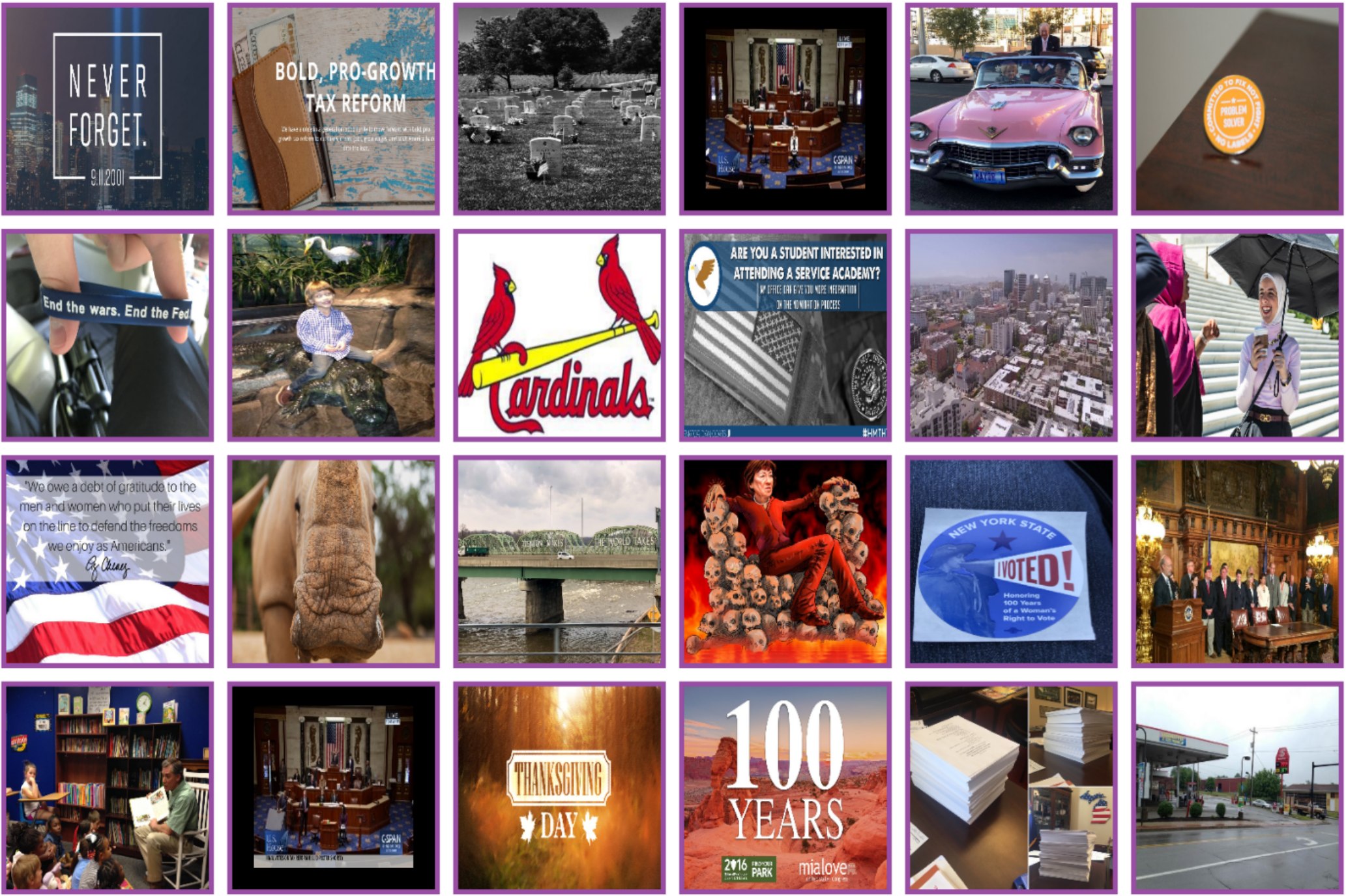}\label{fig:ideological_imagery_types_american}}
    \subcaptionbox{Infographic Imagery}[8.5cm]{\includegraphics[width=0.85\columnwidth]{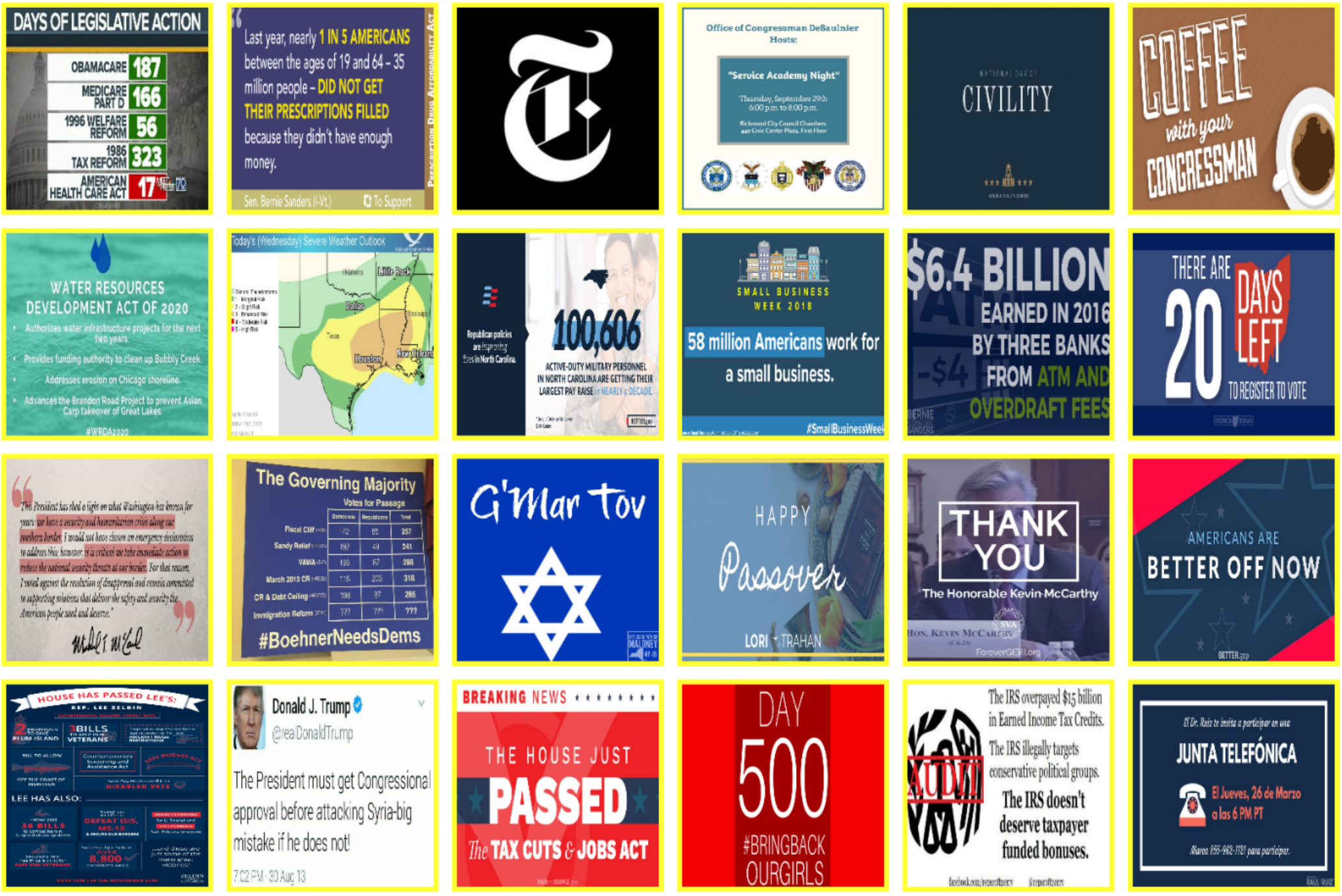}\label{fig:ideological_imagery_types_info}}
    \subcaptionbox{Imagery of People}[8.5cm]{\includegraphics[width=0.85\columnwidth]{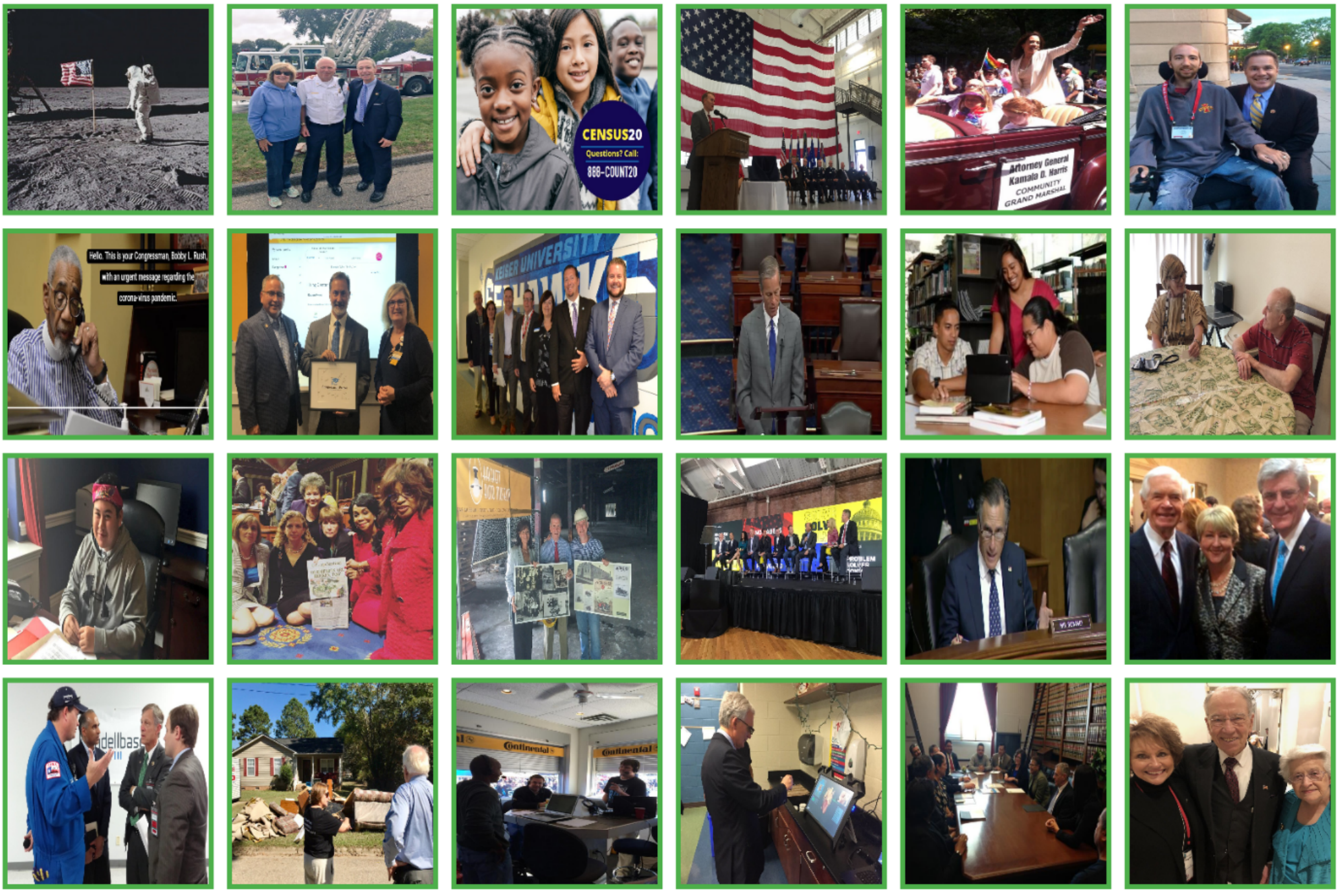}\label{fig:ideological_imagery_types_people}}
    \caption{Four Consistent Types of Images Used By US Politicians.}
    \label{fig:ideological_imagery_types}
\end{figure*}

\subsection{Qualitative Analysis of Imagery}

As discussed in \textbf{RQ1}, we observe that politicians share images from eight different clusters of images, with examples in Figure~\ref{fig:efficientnet_clusters}.
Of these image clusters, some are more strongly correlated with a political ideology than others.
A natural question then concerns what kinds of images are captured by these clusters.

In a post-hoc qualitative assessment of the clusters produced by each embedding model, we find substantial consistency in the kinds of images these clusters represent.
First, we find that each image embedding model contains clusters of document-based imagery, like those in Figure \ref{fig:ideological_imagery_types}a--Cluster 1 in EfficientNetB1, Cluster 5 in ResNet50, Clusters 3/7 in VGG19, Clusters 1/4 in InceptionV3, \hl{and Clusters 0/1/6 in ConvNeXt}. 
Nicely, linear-model coefficients for these embedding models are largely consistent in that document-based images correlate significantly with liberal ideological lean.

Next, the majority of clusterings produce a cluster that contains patriotic imagery, like that shown in Figure \ref{fig:ideological_imagery_types}b, including images of the US flag and other national symbols; Americana, like football games; and national holidays like the Fourth of July.
This American-centric cluster corresponds to Cluster 3 in EfficientNetB1, and Cluster 4 in both ResNet50 and VGG19.
As with clusters of document-images, these Americana clusters are also significantly correlated with conservative imagery across these three embeddings.

Thirdly, infographics are commonly shared by politicians, as shown in Figure \ref{fig:ideological_imagery_types}c.
These images generally present information about political topic, such as who benefits from minimum wage increases or incidence of mass shootings across the US.
These infographics are often in the form of statistics, showing a textual heading and a numeric value, such as the number of jobs in Oklahoma that depend on manufacturing (10,400) or the number of individuals who have signed up for healthcare since the Affordable Care Act's passage (7 million).
EfficientNetB1's Cluster 5 captures these images, as does VGG19's Cluster 7 and ResNet50's Cluster 3, all of which significantly correlate with liberal ideological positions.

Lastly, images of people, especially groups of people are ubiquitous in the data (see Figure \ref{fig:ideological_imagery_types}d), as one might expect given politicians' campaigns and engagements with their constituents, and the type of imagery on which \citet{10.1609/icwsm.v14i1.7338.2020} focuses.
These images come in a variety of styles, from posed photos and gladhanding to more candid shots of groups of various sizes.
Clusters 2, 4, 6, and 7 in EfficientNetB1 capture these groups, with Clusters 2 and 4 primarily comprising candid images, with large/small groups of individuals engaged in some task and not facing the camera, respectively.
EfficientNetB1's Clusters 6 and 7 instead show more posed images of individuals and groups.
As noted above, these clusters of people are difficult for humans to assess, and the limited effect of Clusters 2 and 4 on ideology may be attributable to this difficulty.
For Clusters 6 and 7, however, we do see significant ideological effects, suggesting liberal politicians tend to use posed images of smaller groups, whereas conservative politicians use images of larger groups.
The above descriptions are, however, based on our qualitative descriptions after these images have already been grouped.
Future work should investigate these descriptions more thoroughly.

\subsection{Types of Images and Political Use}

Our results show a differentiation between the types of imagery shared by conservative and liberal politicians: Politicians who share a larger proportion of patriotic or nationalistic imagery tend to be more ideologically conservative, whereas liberal politicians appear to share a larger portion of diverse imagery, including document-based imagery, and infographics.

Conservative politicians' use of patriotic imagery is well documented, as shown in~\cite{Munoz2017}, where an analysis of presidential candidates' Instagram images demonstrates that Donald Trump posted more patriotic symbols than other candidates.
That same study also shows patriotic symbols garnered more engagement from audiences than many other visual themes.
Across the aisle, liberal MoC's use of infographics is consistent with a separate study on Instagram use by MoCs, where \cite{OConnell2018} shows Republican MoCs are significantly less likely to share infographics and text-oriented posts.

These results suggest politicians across the ideological spectrum may use images for the same goal, but how they use images to express this goal vary.
Both \cite{Munoz2017} and \cite{OConnell2018} support this interpretation, as \cite{Munoz2017} shows presidential candidates on both sides used visuals for self-presentation and especially to present themselves in the ``ideal candidate'' frame through depictions showing superior statesmanship.
For conservatives, though, \cite{Munoz2017} and \cite{OConnell2018} find these candidates express this frame through patriotic symbols and visuals of other political elites.
These depictions potentially represent appeals to authority and social dominance respectively, and much work has shown these aspects are strong predictors of conservative ideology, as discussed at length in~\cite{Kugler2014}.
In contrast, liberal politicians may instead express this theme via data-backed claims; e.g., \cite{Miller2010} demonstrates liberal audiences rely more on systematic processing rather than heuristic processing and are more likely to be persuaded by hard data.
Liberal politicians' use of infographic-style and text-laden visuals may then be more persuasive among their audiences.

\subsection{Images Across Ideological Boundaries}

While the above section discusses types of images that are correlated with political ideology, several types of images have no significant correlation.
These clusters may represent ``politically neutral'' types of images, i.e., these clusters may contain images from politicians on both sides of the aisle.
The use of such images -- e.g., marketing graphics, as in cluster 0, or depictions of meeting with constituents, as in cluster 2 -- may simply be a common part of the politician's role as an elected official, regardless of their ideological position.
\cite{OConnell2018} similarly notes little ideological correlation in MoCs use of images showing their constituents.
Cluster 2 in particular presents an interesting case, as MTurk workers exhibit difficulty in assessing similarity therein, as shown in Figure~\ref{fig:similar_imagery_cluster_proportions}.

Alternatively, these images may present ideological signal through semantic meanings rather than visual structure.
Several images in cluster 0 contain political affiliations, representing politicians or icons traditionally associated with either Democrats or Republicans, and many contain captions that present politically oriented messages.
For instance, certain images are praising conservative politicians or showing screenshots from conservative news channels, but other images contain quotes from liberal icons or information about events involving liberal politicians.
Similarly, images from newscasts often look visually similar, but one may be from CNN while the other is from Fox News.
Separating these types of images may be difficult for object-recognition-based frameworks.
In this context, this cluster may balance itself out with an even number of strongly liberal and conservative images.

These observations yield two potential conclusions:
First, certain types of imagery can be shared by any type of politician, crossing ideological boundaries, such as images of politicians engaging in everyday activities or glad-handing.
Second, other types of imagery do not necessarily correlate with a political ideology because they have high visual similarity but entirely different semantic meaning, and the deep learning frameworks we use are unable to capture this semantic variation.
While more research is needed into these types of imagery, one should exercise caution in using these deep-learning models with these types of images.

\subsection{Broader Perspectives and Ethics}

\subsubsection{A Hazard in Studying Social Media Images}
As we mention above, our findings about humans' struggles in assessing similarity for certain kinds of images makes it is difficult to separate potential modeling issues from actual patterns in ideological use by politicians.
While the image-embedding models used herein are common in computer vision, the types of images used in such tasks tend to have only one perceptual basis: an image's \emph{visual} component.
When analyzing imagery shared online, as in social media, variation in type of image (e.g., cartoons, image macros, etc.) and in semantics are equally significant aspects for image understanding.
We therefore highlight the need for caution when directly transferring these computer-vision methods to this space.
To handle this variety of images, methods that consider joint visual-semantic representations are needed.

This consideration is increasingly crucial as the popularity of memetic imagery increases, and adding captions to imagery being shared online becomes easier.
\cite{OConnell2018} in fact demonstrates that many younger politicians increasingly rely on these kinds of images in their political discourse.
Prior work, such as DeViSE~\cite{devise} and MemeSequencer~\cite{Dubey2018}, have sought to provide these joint representations, but limited resources exist for robust pre-trained models, such as those used herein.
Therefore, when applying machine learning models to political content, additional analysis and care are necessary to address these significant confounders.


\subsubsection{Ethical Considerations}

The above work suggests potential weaknesses in image analysis models, especially with respect to the use of textual overlays to embed anti-social content like hate speech in images.
That is, online spaces may have difficulty separating such content from more benign imagery, and having highlighted this weakness, we may have inadvertently alerted malevolent actors to a weakness they could exploit.
Researchers are already working on this problem, but we specifically identify the problems these weaknesses may cause for research into online political discourse.
We hope that by calling attention to these weaknesses and outlining potential mitigations, as we do above, this potential for exploitation will be somewhat mitigated.

\subsubsection{Recommendations for Researchers} 

Throughout our work, a few key takeaways have manifested, and researchers should consider them when analyzing political imagery. 
First, we have identified several distinct types of imagery, some of which present difficulties for humans when assessing similarity--i.e., clusters 2, 4, and 7, as shown in Figure~\ref{fig:similar_imagery_cluster_proportions}.
In these instances, blindly applying pre-trained object-recognition models may prove problematic, as the main axis of meaning in these image types may not be conveyed by the objects present.
On the other hand, MTurk workers and pre-trained models seem to largely agree on image similarity in other types of images--e.g., clusters 3 and 5--suggesting these image types may be more amenable to applications of pre-trained models.
In short, researchers should be cognizant of the \emph{type} of images they are likely to be analyzing and constrain themselves appropriately, as we see with Xi et al. \shortcite{10.1609/icwsm.v14i1.7338.2020}.

Second, researchers should consider the implications of variation in ideology across image types.
If a particular political party or group uses a particular type of image more, as we see with conservatives and patriotic symbolism and with liberals and infographics, this selection has implications for how these groups may use different platforms.
E.g., if a platform's affordance focuses more on photographic imagery than designed images, like infographics or memes, that affordance likely impacts how liberals may use it compared to conservatives.
In such a case, missing out on a certain type of imagery can lead to omitting an entire political group, or how they communicate using imagery. 
Hence, researchers should consider platform norms and affordances in the kinds of images that are popular and how such norms may differentially impact ideological expression.
Twitter is, of course, no exception here, and researchers should likely pilot similar studies in other platforms accordingly.

Third and lastly, we observe that, of the five pre-trained models we have tested in this paper, \hl{four} of them--VGG19, ResNet50, EfficientNetB1, \hl{and ConvNeXt}--present highly correlated and similar results.
In contrast, InceptionV3 appears to deviate substantially from results built from the other models.
A shallow recommendation here might be to avoid InceptionV3, as it appears both deviant and has the lowest $R^2$ value in Table \ref{tab:ols_results}.
As new models \hl{with novel architectures} are released regularly, however, a more valuable researcher recommendation might be to evaluate results across several pre-trained models.
By examining results across multiple such models, the researcher can demonstrate robustness to model selection and get some insight into consistency of results across these models.
For example, our conclusions might be quite different had we used only InceptionV3, whereas they are likely to be similar if we had used one of the other four.

\subsection{Threats to Validity}

While $k=8$ clusters enables a succinct analysis of image types and their political correlations, it is possible that more specific types of imagery may be hidden by this low number of clusters.
For example, we suggest the null result for significance in some clusters above stems from a limitation in the image embedding model, but it could also stem from selecting a cluster count that is too coarse,  artificially grouping these visually similar but semantically distinct images.
To explore this possibility, we have also run our clustering and linear regression models for other values of $k$, namely $k=12$ and $k=20$.
These results align with results for $k=8$, with the new clusters subdividing the $k=8$ clusters, but no unexpected ideological correlations emerge.
This find suggests our results, namely ideological correlations and distributions of imagery across clusters, are robust to cluster-count selection.
Given that our results are consistent across embedding models as well, and no cluster seems to have a dearth of images, and the number of clusters we find is consistent with both \cite{OConnell2018} and \cite{Munoz2017}, we are confident in these results.

We conduct a similar robustness check using a smaller sample of imagery, to assess whether our results resilient to the selection of images. 
This sample contains a maximum of 21 images per account, for a total of 15,054 images. 
We find that similar types of imagery emerge with consistent ideological correlation, albeit weaker due to the reduced sample of imagery. 
This find further corroborates that our results are consistent across different quantities of imagery.

Other threats arise from our sampling strategies from Twitter and in sampling images from our politicians.
First, while our timeframe is broad, covering 2011 to 2021, the Republican Party held the majority in the US House of Representatives throughout this timeframe. 
As~\cite{OConnell2018} suggests, MoCs behave differently when their party is in the majority versus the minority, so it is possible that this timeframe may bias our sample of MoCs toward particular types of images.
As the makeup of Congress changes, future work could test this possibility.

\section{Conclusions}

Results demonstrate that the types of imagery shared in online social spaces can be a strong indicator of one's ideological position -- at least in the context of US congresspeople.
While this finding is consistent with existing literature, we also find that standard deep learning models for image characterization may capture only a portion of this structure.
Across multiple models of image representation, two visually similar images can convey divergent ideological messages.
Despite this divergence, these models result in consistent types of imagery and correlations with ideological positions.
Certain types of imagery, however, appear to have limited to no significant effect related to ideology, and it is difficult to attribute that result to real patterns of behavior among MoCs or to semantic collapse in visually oriented image models.
Additionally, while four of our five image-embedding models produce largely consistent clusters, one model, InceptionV3, diverges strongly from the other three, suggesting different visual interpretations based on model architectures despite similar training datasets.
We therefore highlight the need for special care when applying automated methods to characterize the variety of images used in online political discourse.

\bibliography{bibliography}

\end{document}